# Lorentz force on an electron in a strong plane-wave laser field and the low-frequency limit for ionization

Jarosław H. Bauer *

Katedra Fizyki Teoretycznej Uniwersytetu Łódzkiego,
Ul. Pomorska 149/153, 90-236 Łódź, Poland

A motion of a classical free charge in an electromagnetic plane wave can be found exactly in a fully relativistic case. We have found an approximate non-parameter form of the suitable equations of motion. In a linearly polarized wave, in the simplest frame of reference, the charge moves along the well-known "figure-8" path. We have numerically calculated the Lorentz force acting on the charge as a function of time. In virtue of this, for the low frequency ionization (or detachment) rate, we discuss a manifestation of nondipole and relativistic effects.

---

*Electronic address: bauer@uni.lodz.pl

# I. INTRODUCTION

Let us consider a classical point charge interacting with an arbitrary intense electromagnetic plane-wave field. The charge can move with a relativistic velocity. As we shall demonstrate below, studying such motion is important from the point of view of theories describing ionization (or detachment) in strong laser fields. In Ref. [1] (Sec. 48, p. 134) there are exact solutions to suitable equations of motion in the simplest frame of reference (i.e. in which the charge is at rest on the average). The solutions for a linear polarization and for a circular polarization of the plane wave [1] have been generalized by us recently [2]. Our result for an electron in the laser field of any elliptical polarization is the following (in the present work we use atomic units: $\hbar = e = m_e = 1$, and we substitute explicitly $-1$ for the electronic charge)

$$x = \frac{a^2}{8c\omega\varepsilon^2}\cos\delta \sin 2(\omega t - kx) \equiv x_0 \sin 2(\omega t - kx), \qquad (1a)$$

$$y = \mp \frac{a}{\omega\varepsilon}\sin(\delta/2)\cos(\omega t - kx) \equiv y_0 \cos(\omega t - kx), \qquad (1b)$$

$$z = \frac{a}{\omega\varepsilon}\cos(\delta/2)\sin(\omega t - kx) \equiv z_0 \sin(\omega t - kx). \qquad (1c)$$

In Eqs. (1) we have assumed that the laser field propagates along the $x$ axis, and its wave vector is $k = \omega/c$ ($\omega$ - the laser frequency; $c$ - the velocity of light). $a$ is the amplitude of the vector potential describing the field, $\delta$ is the ellipticity parameter ($\delta \in [0, \pi/2]$; for the linear polarization $\delta = 0$, and for the circular polarization $\delta = \pi/2$), $\varepsilon = \sqrt{c^2 + a^2/2c^2}$, and the signs $\mp$ correspond to two different helicities. The electric field vector ($\vec{E} = -c^{-1}\partial\vec{A}/\partial t$) has the amplitude $E_0 = (a\omega/c)\cos(\delta/2)$. (See Ref. [2] for more details.) In Eqs. (1) we have also defined $x_0, y_0, z_0$ - the amplitudes of motions along the respective axes.

## II. ANALYTICAL RESULTS

Equations (1) are nonlinear, and in general require a numerical treatment to find $x, y, z$ as a function of $t$. However, when the below mentioned condition is satisfied

$$kx_0 \ll 1 , \tag{2}$$

one can expand the right-hand sides of Eqs. (1) in a Taylor series. In further part of this work we assume that Eq. (2) is valid. For any finite $\omega, E_0$ one can easily show that $kx_0 < 1/4$ always. The latter value is for $\omega = const$, $E_0 \to \infty$ or $E_0 = const$, $\omega \to 0$. Neglecting terms of the order of $(kx_0)^2$ and higher, we find the following approximate solutions to Eqs. (1)

$$x(t) = x_0 \sin 2\omega t (1 - 2kx_0 \cos 2\omega t) , \tag{3a}$$

$$y(t) = y_0 \cos \omega t (1 + kx_0 - kx_0 \cos 2\omega t) , \tag{3b}$$

$$z(t) = z_0 \sin \omega t (1 - kx_0 - kx_0 \cos 2\omega t) . \tag{3c}$$

Then one can easily find components of the velocity vector of the electron

$$\dot{x}(t) = 2\omega x_0 (\cos 2\omega t - 2kx_0 \cos 4\omega t) , \tag{4a}$$

$$\dot{y}(t) = -\omega y_0 \sin \omega t (1 - kx_0 - 3kx_0 \cos 2\omega t) , \tag{4b}$$

$$\dot{z}(t) = \omega z_0 \cos \omega t (1 + kx_0 - 3kx_0 \cos 2\omega t) , \tag{4c}$$

and components of its acceleration vector

$$\ddot{x}(t) = -4\omega^2 x_0 \sin 2\omega t (1 - 8kx_0 \cos 2\omega t) , \tag{5a}$$

$$\ddot{y}(t) = -\omega^2 y_0 \cos\omega t (1 + 5kx_0 - 9kx_0 \cos 2\omega t) ,  \tag{5b}$$

$$\ddot{z}(t) = -\omega^2 z_0 \sin\omega t (1 - 5kx_0 - 9kx_0 \cos 2\omega t) . \tag{5c}$$

Equations (5) determine the Lorentz force (coming from both the electric $\vec{E}$ and the magnetic $\vec{B}$ components of the laser field) acting on the electron, as a function of time. To the best of our knowledge, Eqs. (3-5) have been obtained for the first time here, and are the main (analytical) result of this paper. Let us denote a position of the electron (in the simplest frame of reference) as $\vec{r}(t) = (x(t), y(t), z(t))$. Then the Lorentz force (defined as a time-derivative of a relativistic momentum) is given by

$$\vec{F}_{rel}(t) = \frac{1}{\left(1 - \frac{\dot{\vec{r}}(t)^2}{c^2}\right)^{1/2}} \ddot{\vec{r}}(t) + \frac{\dot{\vec{r}}(t) \ddot{\vec{r}}(t)}{c^2 \left(1 - \frac{\dot{\vec{r}}(t)^2}{c^2}\right)^{3/2}} \dot{\vec{r}}(t) . \tag{6}$$

(There is a scalar product of $\dot{\vec{r}}(t)$ and $\ddot{\vec{r}}(t)$ in a numerator of the second term in Eq. (6).) If $|\dot{\vec{r}}(t)| \ll c$, one obtains from Eq. (6) the nonrelativistic approximation to the Lorentz force:

$$\vec{F}_{nonrel}(t) = \ddot{\vec{r}}(t) . \tag{7}$$

The above equation contains Eqs. (5) and takes into account nondipole effects. If one also puts $x_0 = 0$ in Eqs. (3-5), one obtains from Eq. (7) the nonrelativistic Lorentz force in the dipole (or long wavelength) approximation. One usually assumes that the dipole approximation is valid, when $k = \omega/c \ll 1$. However, according to Reiss [3], nondipole effects should appear for the $H(1s)$ atom, if

$$x_0 \gtrsim 1 \tag{8}$$

($x_0$ here is denoted as $\beta_0$ in Ref. [3]). We agree that nondipole effects can appear in angular distributions of photoelectrons from strong-field ionization, if the criterion (8) is obeyed. Nevertheless, if one looks at a total ionization (or detachment) rate, Eq. (8) may be too restrictive, particularly in the low-frequency limit of strong-field ionization, as we shall demonstrate below.

### III. DISCUSSION AND NUMERICAL RESULTS

Let us consider now the linear polarization ($\delta = 0$), which is of most experimental interest. From Eqs. (1) we obtain $y(t) = 0$, and the electron moves in the $xz$ plane. The motion takes place along the "figure-8" path *ABCDAEFGA* (shown schematically in Fig. 1), which is covered every laser cycle. It follows from Eqs. (1) that the ratio $x_0 / z_0$ grows monotonically with increasing a laser field intensity $I$ (for the linear polarization $I = E_0^2$) from 0 (for $I = 0$) to $\sqrt{2}/8 \approx 0.177$ (for $I \to \infty$). For strong laser fields, almost for all $t$, the distance $\sqrt{x(t)^2 + y(t)^2 + z(t)^2}$ (calculated from Eqs. (3), for any elliptical polarization) is much larger than a radius of an atom (or ion). Therefore, total forces acting on the ionized (or detached) electron during its motion in strong laser fields are nearly equal to those of a free motion, because binding forces (Coulomb or short-range) are much weaker. Moreover, according to the quasi-static limit of the ionization theory by Keldysh [4], the electron escapes, when both the $\vec{E}$ and $\vec{B}$ fields are close to their maximum values during the laser cycle. If the laser frequency $\omega$ is much lower than a characteristic atomic frequency, the Keldysh adiabaticity parameter $\gamma$ [4] obeys the condition

$$\gamma = \frac{\omega \sqrt{2 E_B}}{E_0} \ll 1 \ . \tag{9}$$

$E_B$ denotes here a binding energy of the atom or ion. In further part of this work we assume that the inequality (9) is satisfied. In the limit $\omega \to 0$ (then also $\gamma \to 0$, if $E_0 = const$) the ionization rate $\Gamma$ is approximately given by an expression of the type

$$\Gamma \approx f(E_0)\exp(-C/E_0), \qquad (10)$$

where $C > 0$ is a constant (or nearly a constant), $f(E_0)$ is a relatively slowly-varying function of $E_0$, and $\exp(-C/E_0)$ grows rapidly with $E_0$. (Both $f(E_0)$ and $C$ depend also on $E_B$ and the initial-state wave function.) In Keldysh's theory (see Eq. (20) of Ref. [4]) the pre-exponential factor $f(E_0)$ is not the same as in the static-field theories [5-7], when the ionization rate is averaged over the cycle of the electric field ($E(t) = E_0 \sin\omega t$). However, the exponential factor $\exp(-C/E_0)$ remains the same. One should stress that the well-known dependence (10) is typical not only for the early tunneling theories [4-13], where one usually assumes that $E_0 \ll E_{BSI}$ (BSI denotes the barrier-suppression ionization). Dörr et al. [14] investigated the static-field limit in multiphoton ionization with the help of the Floquet method. For the linear polarization they found that Eq. (10) describes the ionization rate, but does not account for intermediate resonances (which occur for some specific values of $\omega$ and $E_0$). Ilkov et al. [15] confirmed experimentally, that Eq. (10) is approximately valid for $\gamma < 0.5$ and $E_0 < E_{BSI}$. Buerke and Meyerhofer [16] confirmed experimentally a high accuracy of the semiclassical approach [8-13] in the tunneling regime. Scrinzi et al. [17] exactly calculated the static-field ionization rate $\Gamma_{stat}$ up to $E_0 = 1$ a.u. for the $H(1s)$ atom. When this $\Gamma_{stat}$ is averaged over the cycle of the electric field ($E(t) = E_0 \sin\omega t$), one numerically obtains the ionization rate $\Gamma_{stat}^{av}$, which is really of the type of Eq. (10). For $0.03$ a.u. $\leq E_0 \leq 1$ a.u. $\Gamma_{stat}^{av}$ is equal to the averaged ionization rate of Landau [6] times a factor of the order of $0.1 \div 1$, as shown in Fig. 2 of Ref. [18]. At the same time, for $0.03$ a.u. $\leq E_0 \leq 1$ a.u., $\Gamma_{stat}^{av}$ changes over about ten orders of magnitude. An approximate empirical formula (the so-called TBI formula) for the static-field ionization rates for atoms and molecules and fields up to $E_0 \geq \sim E_{BSI}$ was found by Tong and Lin [19]. The TBI formula can also be treated as of the type of Eq. (10), with a slowly-varying function $f(E_0)$. In Refs. [4-19] the nonrelativistic and dipole approximation was applied to a description of the ionization, but in other works [20-24] the magnetic-field component or relativistic effects of the laser field were taken into account. In the limit $\omega \to 0$, in all the cases [4-24], Eq. (10) is approximately valid.

In Eq. (10), the ionization rate $\Gamma$ depends strongly on $E_0$, which is the laser field parameter. The Coulomb (or short-range) force, acting on the electron, is present in Eq. (10) only through constants included in $f(E_0)$ and $C$. Therefore, for a given initial state of the atom (or ion), the ionization rate $\Gamma$ is determined by the amplitude of the electric field vector $E_0$. In the nonrelativistic and dipole approximations, $E_0$ is equal (in atomic units) to the maximal Lorentz force exerted on the electron during its motion along the "figure-8" path (which simply becomes a line segment in this case). The force (in the simplest frame of reference) can be calculated from Eqs. (5) and (7). In the intermediate range of the laser field parameters one can keep the nonrelativistic theory, but one has to take into account the magnetic-field component of the laser. Then both the electric field $\vec{E}$ and the magnetic field $\vec{B}$ depend only on time, and $\vec{B} = \hat{n} \times \vec{E}$ ($\hat{n}$ is a unit vector in the propagation direction). In the fully relativistic case, one should replace Eq. (7) with Eq. (6) to calculate the Lorentz force acting on the electron.

Taking into account our discussion related to Eq. (10), one can suppose that in the limit $\omega \to 0$, during the motion shown in Fig. 1, the electron is most probably emitted near the points $C$ and $F$ (i.e. when both the $\vec{E}$ and $\vec{B}$ fields are close to their maxima). The tunneling picture of ionization in static electric fields [5-7] suggests that the electron is emitted mostly in the direction of the electric field vector. Fig. 1 indicates that the electron may be emitted not only in the polarization direction (in the simplest frame of reference), but also at some little angle in relation to this direction. This happens, when the electron is not emitted exactly from the points $C$ or $F$ (note that for the circular polarization the electron always escapes in the polarization plane, in the simplest frame of reference). The condition (8) is important, if one looks at angular distributions of photoelectrons. Indeed, such effects were theoretically predicted (usually with the exponential accuracy) for different polarizations in relativistic (or at least nondipole) strong-field photoionization (see, for example, Refs. [25-29]). As it is generally known, the classical free point charge in the monochromatic plane-wave electromagnetic field moves with the so-called drift velocity, which is constant and parallel to the wave vector $\vec{k} = k\hat{n}$ [30,31]. As a result, the ionized electron has a greater momentum in the forward $\hat{n}$-direction than it would have in the dipole approximation, in which the drift velocity is zero. In the nonrelativistic approximation, the average drift per cycle in the propagation direction is of the order of $E_0^2/c\omega^3$. According to Joachain *et al.* [30], nondipole effects could appear, if the above mentioned drift would be at least equal to 1 *a.u.* (for the $H(1s)$ atom). The latter condition is equivalent to Eq. (8) (up to a constant factor of the order of unity). However, binding forces (Coulomb or short-range) make the electronic trajectory more complicated [32], and the magnetically induced drift may be overcome by the attraction of the nucleus [33].

In our opinion, Eq. (10) suggests that nondipole or relativistic effects could appear in the ionization rate, if they would appear in the Lorentz force, i.e. when Eq. (6) would differ significantly from Eq. (7). In Figs. 2 and 3 we have investigated the Lorentz force as a function of time during the motion along the first half of the "figure-8" path. The solid lines show fully relativistic results, based on Eqs. (4-6). The dashed lines show the nonrelativistic and nondipole results, based on Eqs. (4,5,7), and the dotted lines show the nonrelativistic results in the dipole approximation (Eqs. (4,5,7) with $x_0 = 0$). Each figure contains numerical values of some essential parameters, namely $\gamma$ (with $E_B = 0.5$ *a.u.*, for the $H(1s)$ atom), $z_f = 2U_P/c^2$ ($U_P$ - the ponderomotive potential; see also Refs. [2,3]), $x_0$, and $kx_0$ (the latter value is shown to confirm validity of Eq. (2) in each case). In Figs. 2(a)-(c) $\omega = 0.0043$ *a.u.*, what corresponds to $CO_2$ laser radiation ($\lambda = 10.6$ $\mu m$), and in Figs. 3(a)-(c) $\omega = 0.057$ *a.u.*, what corresponds to $Ti:sapphire$ laser radiation ($\lambda = 800$ *nm*). The Keldysh parameter $\gamma = 0.1, 0.033, 0.01$ at the consecutive plots (a), (b), and (c). In Figs. 2(a) and 3(a) the three above mentioned Lorentz forces, are nearly indistinguishable from each other, in spite of quite large values of $x_0$. In the next figures, (b) and (c), the amplitude of the electronic motion in the propagation direction ($x_0$) grows significantly. In Figs. 2(b) and 3(b) we show that the nonrelativistic dipole approximation becomes insufficient for $x_0 = 190$ *a.u.* (if $\omega = 0.0043$ *a.u.*), and for $x_0 = 14$ *a.u.* (if $\omega = 0.057$ *a.u.*). As one should expect, for extremely intense fields, the relativistic description is necessary (see Figs. 2(c) and 3(c)). However, it follows from Figs. 2(b),(c) and 3(b),(c) that the nonrelativistic nondipole approximation for the Lorentz force works much better than the nonrelativistic dipole approximation. Furthermore, the former one becomes the most accurate for $t = 0.25$ (in laser cycles), when the ionization is the fastest. There is a simple explanation of this fact, namely, near $t = 0.25$ (and $t = 0.75$), the velocity of the electron $v$ achieves a local minimum during its "figure-8" motion. (This corresponds to the points $C$ and $F$ in Fig. 1). For example, in Figs. 2 and 3, for $t = 0.25$, one has $v/c = 0.0013, \ 0.012, \ 0.12$ for Figs. (a), (b), and (c) respectively. It appears that even for extremely strong fields, the electron mostly moves with the nonrelativistic ($v \ll c$) velocity, when it is ionized.

## IV. REMARKS AND CONCLUSIONS

Our present results throw some light on a recent experiment [34] with argon and pulsed-laser 800-nm radiation of the linear polarization at intensities of up to $10^{19} \ W/cm^2$. In Ref. [34] Chowdhury *et al*. confirmed validity of the well-known ADK/WKB tunneling model [8-13] (which is the nonrelativistic and dipole approximation theory) for the average Keldysh parameter $\gamma \approx 0.03$. According to the authors of Ref. [34], the results of their experiment may be interpreted within a two-step model, where the initial tunneling ionization process is dominated by the nonrelativistic effects, while the photoelectron continuum dynamics are strongly relativistic.

In conclusion, we have derived a non-parameter form of the classical relativistic equations of motion for the electron in the monochromatic plane-wave laser field (of an arbitrary intensity and ellipticity). Our approximate solution (in the simplest frame of reference) is based on Eq. (2), which is valid even for superstrong laser fields. We have numerically calculated the Lorentz force acting on the ionized electron during its "figure-8" motion (as a function of time) in the linearly polarized field for two frequencies of an experimental interest. In the low-frequency limit, a manifestation of the nonrelativistic and relativistic effects in the Lorentz force may be a quite good indication of such effects in the ionization rate. For higher frequencies, the simple "tunneling-like" formula (10) is not valid. However, it is quite likely that our argument, based on the above mentioned effects in the Lorentz force, is of some importance.

ACKNOWLEDGMENTS

The author is indebted to Professor Piotr Kosiński for interesting discussions related to this work and useful remarks on a previous version of the manuscript. The present paper has been supported by the University of Łódź.

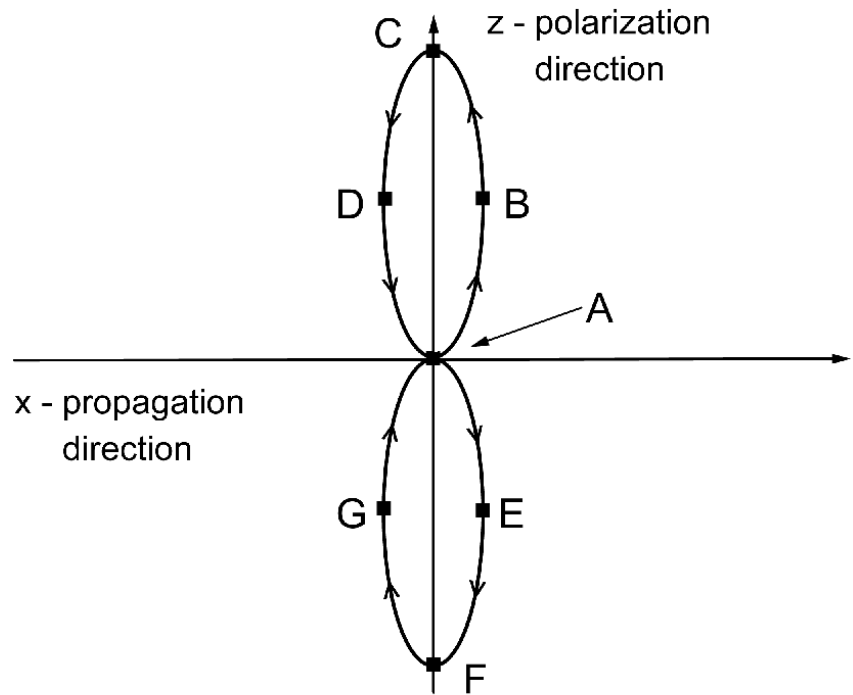

FIG. 1. Motion of the charge along the "figure-8" path (shown schematically) in the linearly polarized plane-wave laser field (in the simplest frame of reference, in the fully relativistic case; see the text for more details).

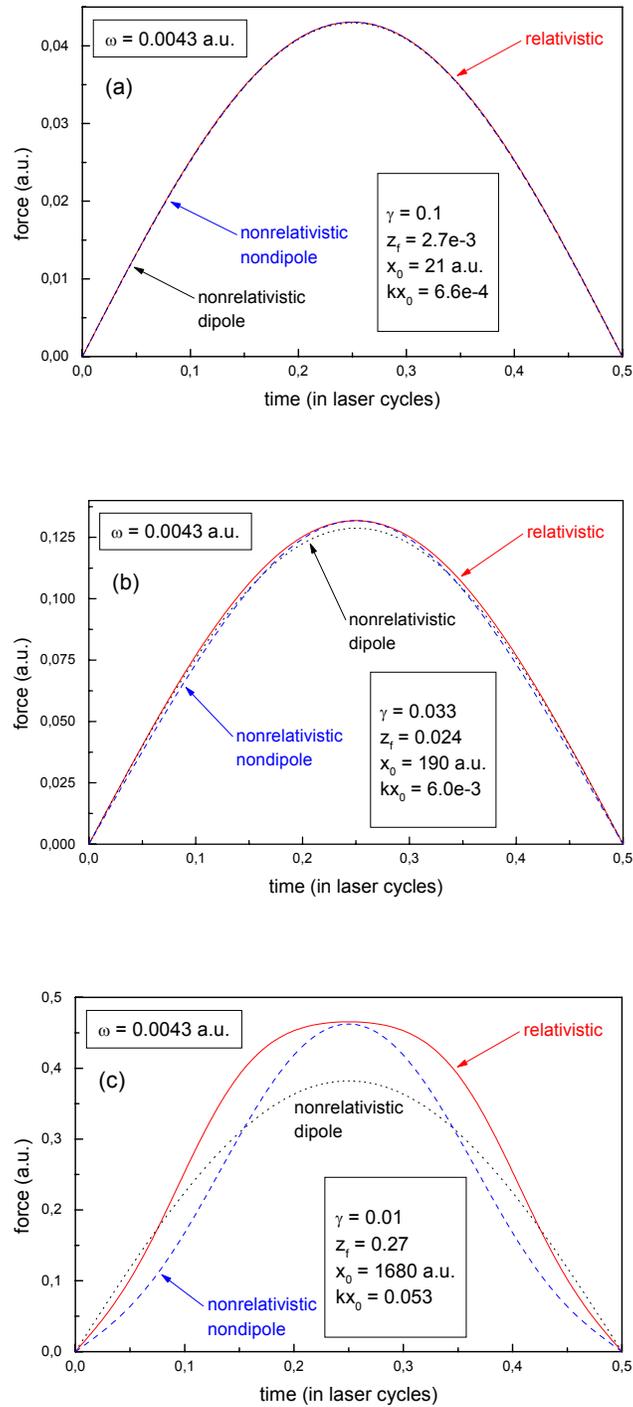

FIG. 2 (color online). The Lorentz force acting on the electron as a function of time during the first half of the laser cycle, for $\omega = 0.0043$ $a.u.$ Three values of the Keldysh parameter are fixed here ($\gamma = 0.1$, $0.033$, $0.01$ in the plots (a), (b) and (c) respectively). The laser intensity increases from (a) to (c). ($2.7e-3$ denotes $2.7 \cdot 10^{-3}$; see the text for more details.)

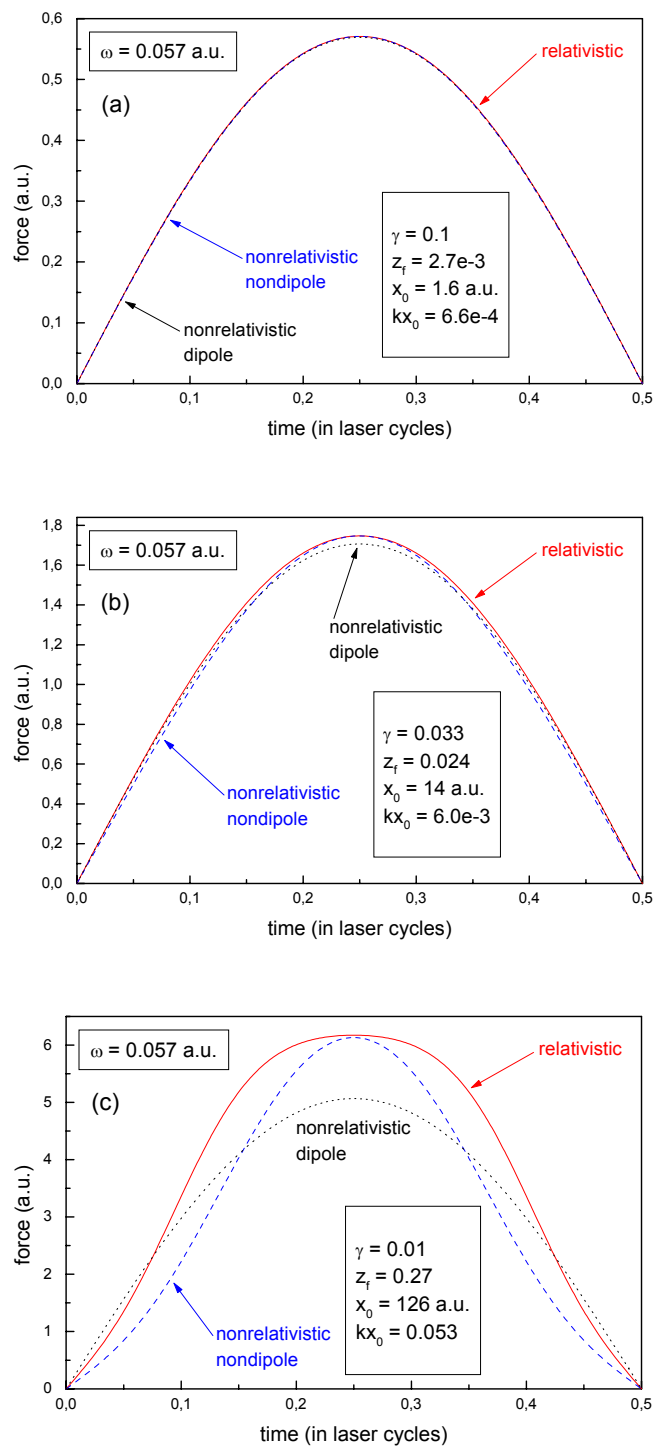

FIG. 3 (color online). The same as in Fig. 2, but for $\omega = 0.057$ *a.u.*